\documentclass[12pt]{article}
\usepackage{graphicx}
\usepackage{times}
\usepackage{hyperref}

\newcommand{\beq}{\begin{equation}}
\newcommand\hiddenref[1]{\sbox0{\cite{#1}}}

\def\cE{ \mathcal{E} }

\newcommand{\bu}{{\bf u}} 
 
\newcommand{\bF}{{\bf f}}

\newcommand{\kL}{{k_{_L}}}
\newcommand{\kH}{{k_{_H}}}
\newcommand{\kB}{{k_{_B}}}

\newcommand{\grad}{\nabla  }

\usepackage{ulem}
\usepackage{ulem}

\def\SOUT#1{}

\usepackage{amsmath}

\newenvironment{sciabstract}{%
\begin{quote} \bf}
{\end{quote}}

\begin{document} 

\title{Large-scale self-organisation in dry turbulent atmospheres}



\author{Alexandros Alexakis$^{1}$, 
        Raffaele Marino$^{2}$,
        Pablo D.~Mininni$^{3}$, \\
        Adrian van Kan$^{4}$,
        Raffaello Foldes$^{2}$,
        Fabio Feraco$^{2,5}$
\\
\normalsize{$^{1\ast}$
LPENS, ENS, Université PSL, }\\
\normalsize{CNRS, Sorbonne Université, Université Paris-Diderot}
\\
\normalsize{$^{2}$CNRS, \'Ecole Centrale de Lyon,  INSA de Lyon, Universit\'e Claude Bernard Lyon 1,} \\ 
\normalsize{ LMFA, UMR5509, F-69134 Écully, France}
\\
\normalsize{$^{3}$Universidad de Buenos Aires, Departamento de Física, }  \\
\normalsize{(INFINA), Ciudad Universitaria, 1428 Buenos Aires, Argentina}
\\
\normalsize{$^{4}$ Department of Physics, University of California, Berkeley, }\\ 
\normalsize{ Berkeley, 94720, California, USA} \\
\normalsize{$^{5}${Leibniz-Institute of Atmospheric Physics, University of Rostock}, }\\ \normalsize{{Schloßstraße 6}, {K\"uhlungsborn}, {18225}, {Germany}} \\
\normalsize{$^\ast$ E-mail:  alexakis@phys.ens.fr}}


\date{}




\baselineskip24pt


\maketitle


\begin{sciabstract}

This work has been published in:

Science 383, 1005--1009 (2024). \,\, DOI:10.1126/science.adg8269

\

How turbulent convective fluctuations organise to form large-scale structures in planetary atmospheres remains a question that eludes quantitative answers.
The assumption that this process is the result of an inverse cascade was suggested half a century ago in two-dimensional fluids, but its applicability to atmospheric and oceanic flows remains heavily debated, hampering our understanding of the energy balance in planetary systems.
We show with direct numerical simulations of spatial resolutions of $12288^2\times 384$ points that rotating and stratified flows can support a bi-directional cascade of energy, in three dimensions, with a ratio of Rossby to Froude numbers comparable to that of the Earth’s atmosphere.
Our results establish that in dry atmospheres spontaneous order can arise via an inverse cascade to the largest spatial scales.
\end{sciabstract}

\section{Introduction}\label{sec1}


Flow structures thousands of kilometers wide are not uncommon in the atmosphere of the Earth and of other planets. 
The energy of these structures could originate from processes associated with the global atmospheric circulation, but also from smaller scale convective turbulence. 
In the latter case, small-scale eddies conspire to self-organise into larger structures. 
Such a process  goes against our daily life experience where turbulence generates smaller-scale erratic structures 
%
as one observes when pouring milk in a coffee cup.
One thus needs to come up with convincing mechanisms for how such large-scale organization can take place in planetary atmospheres. 

One of the most important theoretical discoveries in the twentieth century in the field of non-equilibrium physics is the phenomenon of self-organization, which spontaneously creates large-scale order out of small-scale disorder. 
One of the first examples of this process was given by Onsager with the statistical mechanics of a gas of point vortices \cite{Onsager1949} that was later generalized to two-dimensional turbulent flows 
\cite{Kraichnan1967, Kraichnan1980, leith1968diffusion, batchelor1969computation}.
A two-dimensional flow conserves an additional invariant, the enstrophy, given by the mean squared vorticity.
The relation between energy and enstrophy leads to an incompatibility for the simultaneous 
bulk transfer of both quantities to the small scales. As a result vortices self-interact transferring  enstrophy to smaller scales while energy is transferred to larger scales.  
This process takes place  on a continuum of scales, forming a constant flux of energy from small to large scales in what is known as an inverse energy cascade, as opposed to the disordered forward energy cascade observed in three-dimensional turbulence that is directed  to small scales.

 Although planetary atmospheres are often very thin (the Earth's atmosphere has horizontal synoptic scales of the order of 1000 km, and a pressure scale height of 15 km), the corresponding flows are far from being two-dimensional.  
Nonetheless, two-dimensionality is not imperative for the appearance of self-organisation. 
Three-dimensional rotating and stratified flows (two key ingredients of atmospheric dynamics) 
conserve a different invariant, the potential vorticity, that  can also lead to an inverse cascade.
This happens in the quasi-geostrophic limit where rotation and stratification are asymptotically strong \cite{charney1971geostrophic}.
In this regime the flow is restricted to modes that satisfy hydrostatic balance, and as a result gravito-inertial waves are filtered out. 
Inverse cascades can also be present in rotating Rayleigh-B\'enard convection \cite{Roberts_1968, Busse_1970,rubio2014upscale, favier2014inverse, Siegelman_2022}, where in this case a generalized quasi-geostrophy limit can be considered that partially preserves gravito-inertial modes.
However for most planetary flows, the quasi-geostrophic limit is at best a crude approximation, with gravito-inertial waves composing a significant part of the energy budget 
cascading energy forward 
\cite{vanneste2004exponentially,vanneste2013balance,thomas2021forward,thomas2020turbulent}.
Thus, an inverse cascade in planetary atmospheres caused either by two-dimensionality or quasi-geostrophy remains conjectural.

Could atmospheric dynamics display an inverse cascade away from these limits? 
In recent years it has been demonstrated that 
a hybrid state can be reached such that large scales cascade energy inversely while smaller scales cascade energy forward in what is now known as a bidirectional 
cascade \cite{Alexakis_2018}.
Bidirectional cascades have been observed early on with direct numerical simulations (DNS) \cite{smith1996crossover,Celani_2010,seshasayanan2014edge,benavides2017critical,sozza2015dimensional,van2020critical}.
In rotating and stratified flows, simulations also indicate the presence of bidirectional cascades \cite{marino2013inverse,pouquet2013geophysical,Marino2015},
though in a regime where rotation and stratification are comparable in strength, which is typical for the ocean but not for the atmosphere.

Nonetheless, the existence of self-organisation processes through a bidirectional cascade in planetary atmospheres becomes a compelling possibility as recent research using satellite images with cloud tracking analysis, and in-situ aircraft measurements, estimated the flux (and thus also the direction) of the energy cascades in planetary flows in the Earth's atmosphere \cite{byrne2013height,king2015upscale}, the ocean \cite{balwada_22}, and the Jovian atmosphere 
\cite{young2017forward,Siegelman_2022}.
These studies affirmed the presence of both inverse and forward energy cascades depending on the scale examined or on the altitude. 
However, satellite images constrain the measurements to two-dimensional slices ignoring thus any processes occurring in the third direction. 


Up to today there is no definite evidence of whether planetary atmospheric flows satisfy the necessary conditions for a bidirectional cascade to establish itself.
The difficulty in answering such questions lies on the one hand in the fact that information from satellite images is limited,
and on the other hand in the extreme parameter values that are met in planetary atmospheres which are hard to obtain in DNS. 
However, not only the technology to perform high-cadence high-resolution observations of the atmosphere has just started to come along, but also the computational power to perform DNS  of stratified atmospheres in a realistic parameter space has become available. 
Here, with the use of DNS  in a large grid using $40{,}000{,}000$ CPU hours, we establish that  the fluid model of a rotating and stably stratified  dry atmosphere described by the non-hydrostatic Boussinesq equations can generate a bidirectional cascade leading to large-scale organisation of the flow.

\section{Results}\label{sec2}        

We consider a fluid in a Cartesian, triply periodic domain of vertical height $H$ and horizontal dimension $L=32H$, in the presence of gravity, a stable mean density gradient and solid body rotation in the vertical direction \cite{sm}. The dynamics of the system are described by the incompressible velocity field $\bu$ and the normalized density variation $\phi$, governed by the Boussinesq equations \cite{pedlosky1987geophysical,vallis2017atmospheric}:
\begin{eqnarray}
\partial_t \bu + \bu \cdot \grad \bu+2\Omega\times \bu  &=&  -\nabla P 
- {\bf e}_z N \phi +\nu \nabla^2 \bu + \bF , \label{eq:BS1}\\
\partial_t \phi + \bu \cdot \grad \phi &=& N {\bf e}_z\cdot \bu + \kappa \nabla^2 \phi , \label{eq:BS2}
\end{eqnarray}
where $\Omega$ is the solid body rotation rate, $N$ the Brunt-V\"ais\"al\"a frequency, $P$ is the pressure, $\nu$ the viscosity, $\kappa$ the density diffusivity,
 and $\bF$ an external forcing acting at scales $\ell_{_F}\sim H$ injecting energy at a rate $\epsilon$. 
Although this model has some strong simplifications, like periodicity or a simplified forcing mechanism, it is the most elementary model capturing the necessary physics to reproduce atmospheric dynamics.

This system has four independent non-dimensional control parameters:
 (i) the Reynolds number  $\textrm{Re}_\epsilon=\epsilon^{1/3}k_{_H}^{-4/3}/\nu$, 
 (ii) the Prandtl number $\textrm{Pr}=\nu/\kappa$, that here is set to unity, 
 (iii) the Rossby number,  $\textrm{Ro}_\epsilon =\epsilon^{1/3} k_{_H}^{2/3} /\Omega$, 
 and (iv) the Froude number $\textrm{Fr}_\epsilon =\epsilon^{1/3} k_{_H}^{2/3} /N$ (with $k_H=2\pi/H)$. 
We can also define dimensionless parameters based on the domain size $L$ and the flow r.m.s. velocity $U$, as, e.g., $\textrm{Re}=UL/\nu$, $\textrm{Ro}=U/(H\Omega)$, and $\textrm{Fr}=U/(NH)$, which are closer to the definitions used in atmospheric measurements. 

Simulations were  performed at resolutions of $12288^2\times 384$ grid points \cite{Mininni_2011,sm}. As a reference, for $H=15$ km this corresponds to a domain length of $480$ km, and a vertical and horizontal resolution of $39$ m. 
Our simulations are characterised by $\textrm{Re}_\epsilon = 2000$, $\textrm{Ro}_\epsilon = 1$, and $\textrm{Fr}_\epsilon = 0.025$, or alternatively $\textrm{Re} \approx 2\times 10^6$, $\textrm{Ro} \approx 0.4$, and $\textrm{Fr} \approx 0.01$. 
These values are compatible, e.g., with that of the Mesosphere-Lower Thermosphere (MLT) \cite{hanli}. 

\begin{figure*}[h]                                   
\centering
\includegraphics[width=0.9\textwidth]{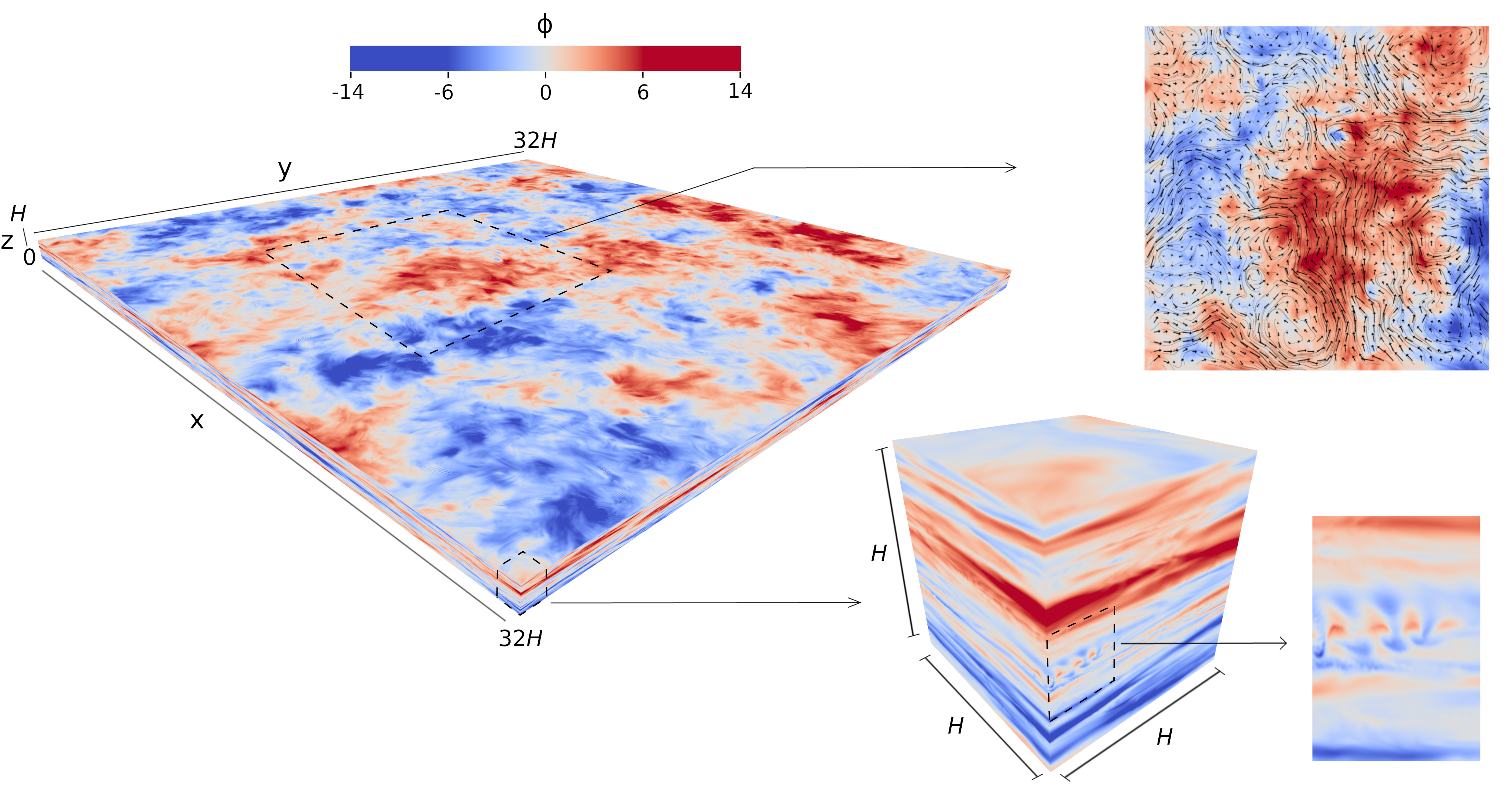}
\caption{{\bf Visualisation of density fluctuations $\phi$ and of the velocity field in the computational domain.} Structures at scales much larger than the forcing (i.e., at the scale of the domain height, with wavenumber $k_f=2\pi/H$) are abundant in the visible horizontal plane (left), indicative of an efficient transfer of the energy towards the lowest modes along the perpendicular direction in Fourier space. The large-scale patterns are visible in the flow visualisation shown by arrows in a zoom in the top right. At the same time, three-dimensional instability patterns and small-scale features are detectable in both horizontal and vertical cuts of the zoomed simulation domain (bottom right), suggesting the action of a forward turbulent cascade. See the Sumplementary Information for a movie of the density fluctuations in the entire domain. Visualisations were done with VAPOR \cite{Clyne_2007}.}
\label{fig1}
\end{figure*}                                         

\hiddenref{Clyne_2007}
Fig. 1 
shows visualisations of the flow and the density field. 
Large-scale structures of horizontal width ten times larger than  $H$ can already be seen by visual inspection.  At the same time, looking at the zoomed-in cross-sections one realizes that these structures are far from being two-dimensional. 
In the large scales, pancake structures of alternating sign of $\phi$ along the vertical, and the emergence of macroscopic cyclones and anti-cyclones, are visible in Fig.~1 (top right panel). These features are observed in weather maps
and are a landmark of the large-scale energy-containing structures in the Earth's atmosphere. 
At the same time small-scale overturning events can be seen in the zoomed in section that are ten times smaller than $H$ (see the bottom right panel in Fig.~1), and are 
also detectable sometimes in the sky as Kelvin-Helmholtz billows. Thus, already at this qualitative level one can testify
the presence of a bidirectional cascade.

To become more quantitative we note that
the inviscid Boussinesq equations conserve the total energy $\cE_T$
given by the sum of
kinetic energy $\cE_K$ and potential energy $\cE_P$. 
Alternatively  we decomposed it into the energy of gravito-inertial modes $\cE_{GW}$ and energy of quasi-geostrophic modes $\cE_{QG}$ where $\cE_T=\cE_{GW}+\cE_{QG}$.
Gravito-inertial modes are dispersive waves-modes due to the combined restoring force of gravity and Coriolis,
while quasi-geostrophic modes are modes that satisfy hydrostatic balance (see Methods for their exact definitions).
These energies are distributed differently in the Fourier space, among the wavenumbers $k_\parallel$, 
identifying horizontal planes, 
and $k_\perp$, 
identifying vertically axisymmetric cylinders. 
 We define three different energy spectra averaged over fixed $k_\parallel$, $k_\perp$, and $k=\sqrt{k_\parallel^2+k_\perp^2}$. We do not define a new symbol for each spectrum, but distinguish between them by their argument, i.e., $E_i(k)$, $E_i(k_\perp)$, and $E_i(k_\parallel)$ where $i$ is $T$, $K$, $P$, $GW$, or $QG$, which stand  for total, kinetic, potential, gravito-inertia wave, and quasi-geostrophic respectively.
In addition we define $E_i(k_\perp,k_\parallel)$ that shows the spectral energy density for a given pair $k_\perp,k_\parallel$.

\begin{figure*}[!h]                                   
\centering
\includegraphics[width=0.475\columnwidth]{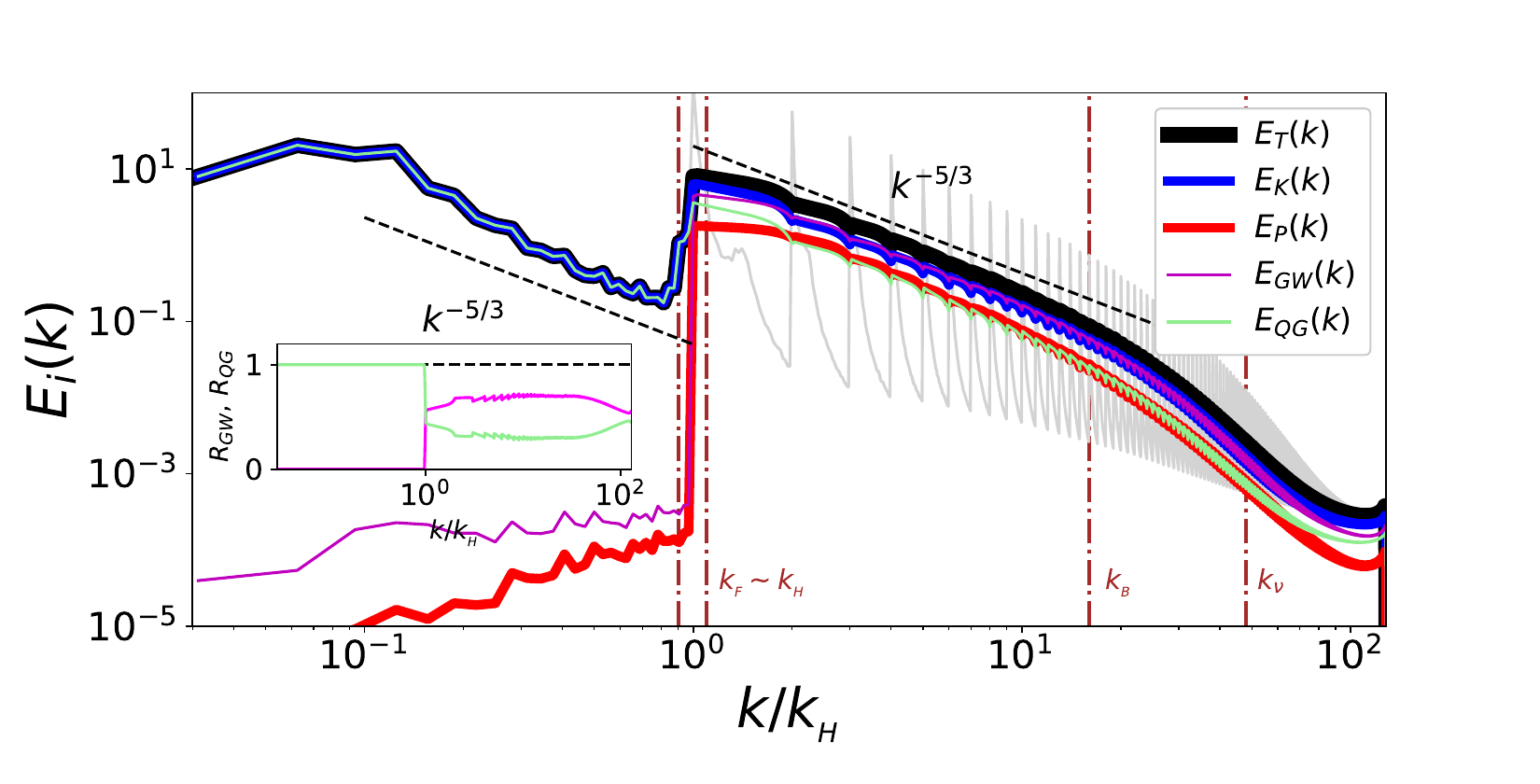}      \includegraphics[width=0.475\columnwidth]{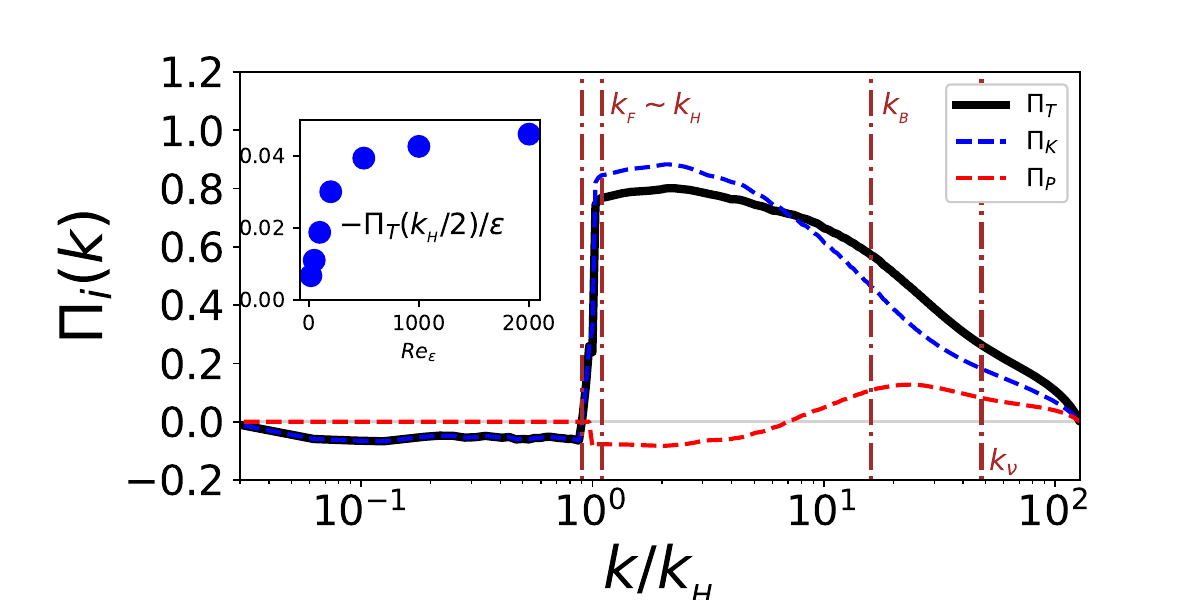} 
\includegraphics[width=0.475\columnwidth]{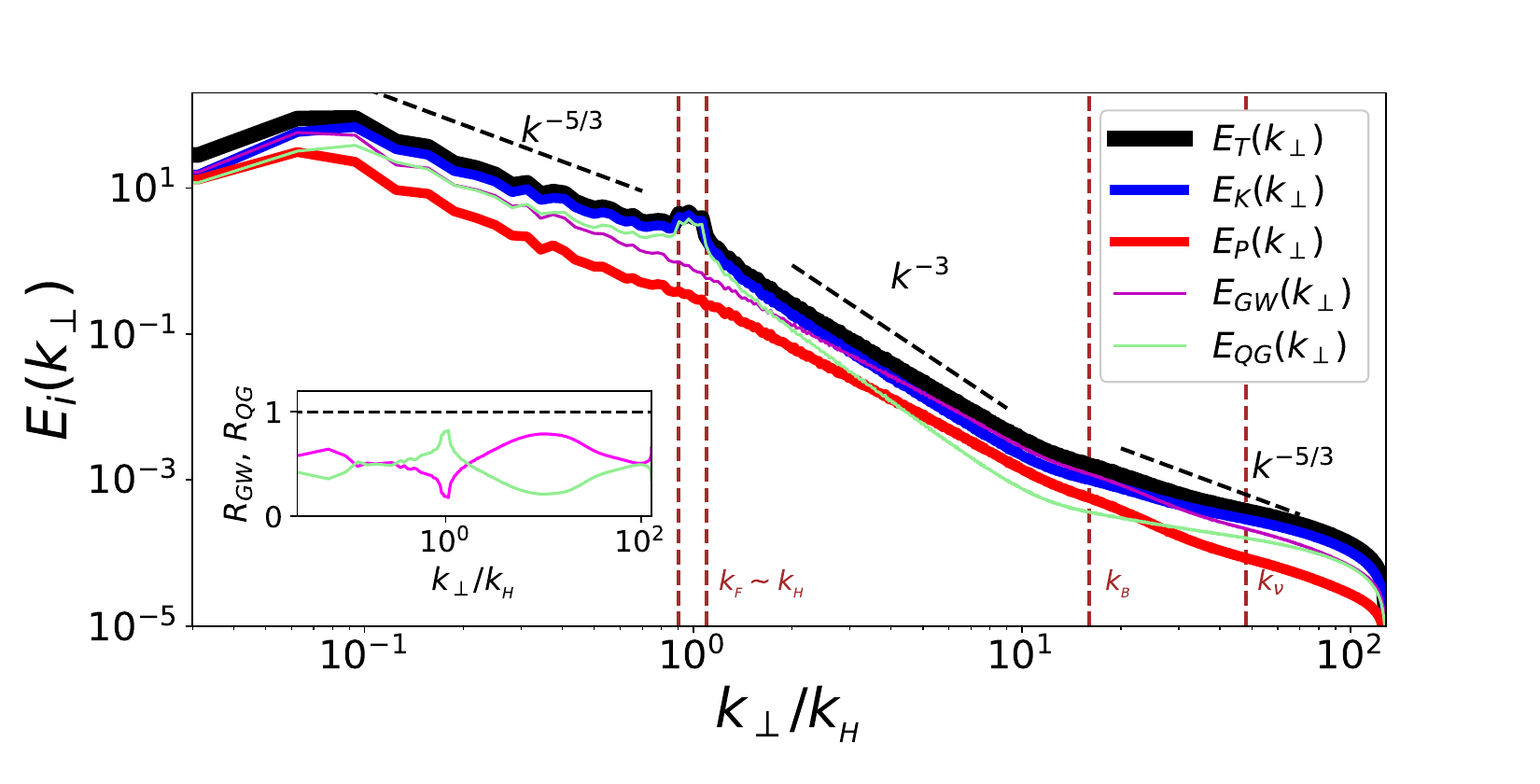}     \includegraphics[width=0.475\columnwidth]{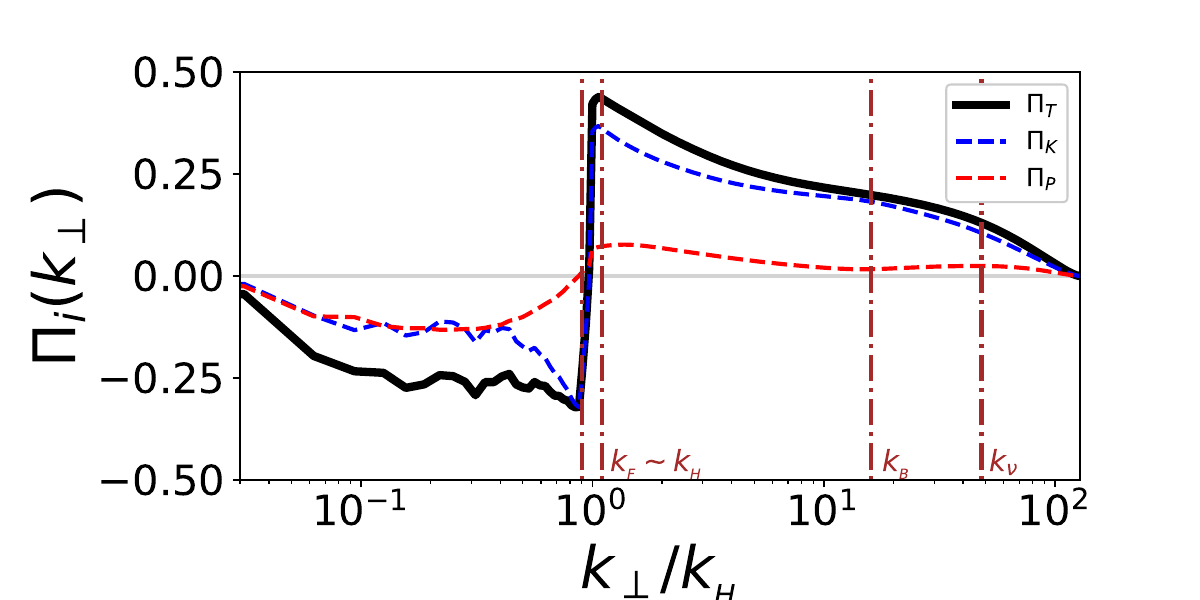} 
\includegraphics[width=0.475\columnwidth]{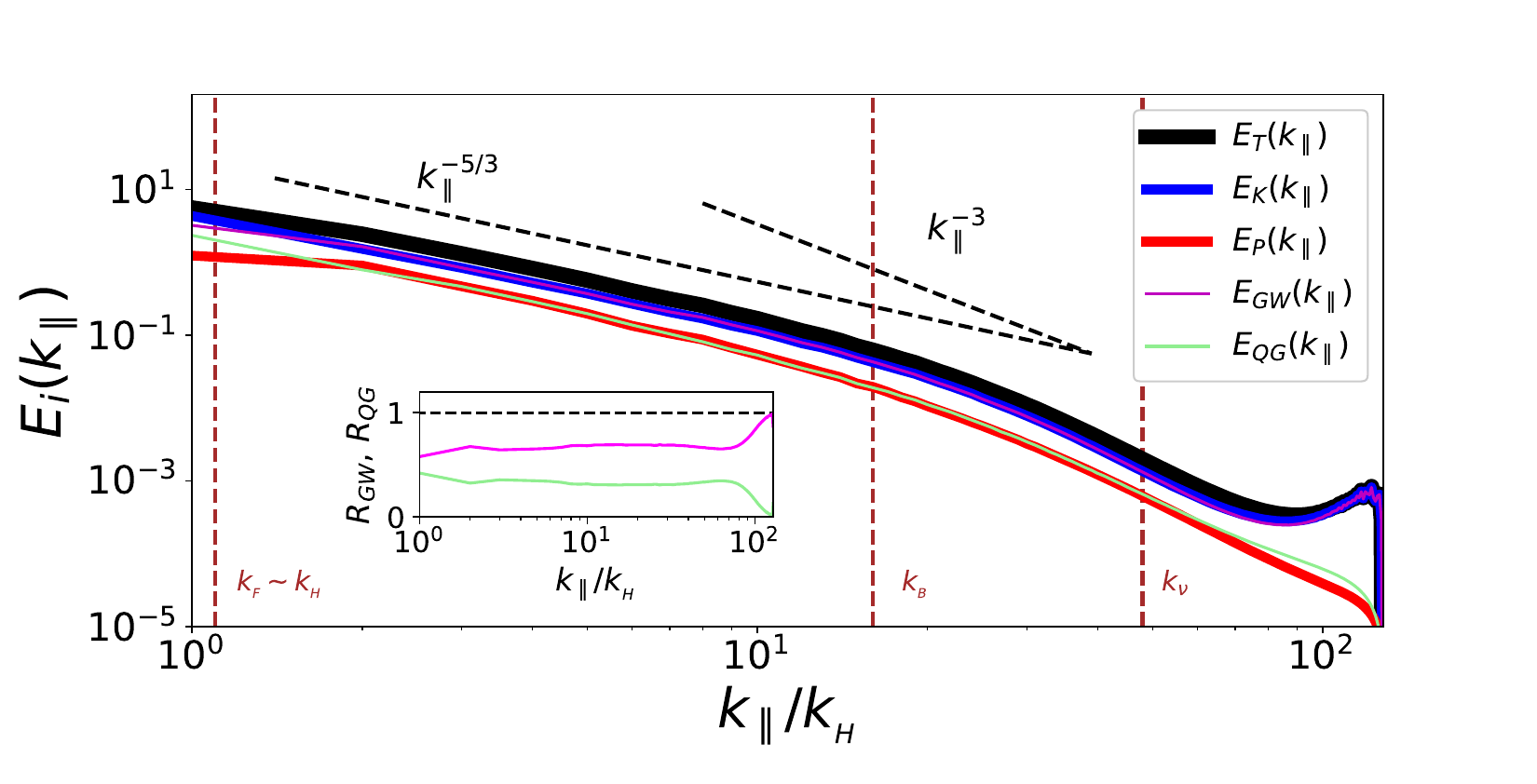} \includegraphics[width=0.475\columnwidth]{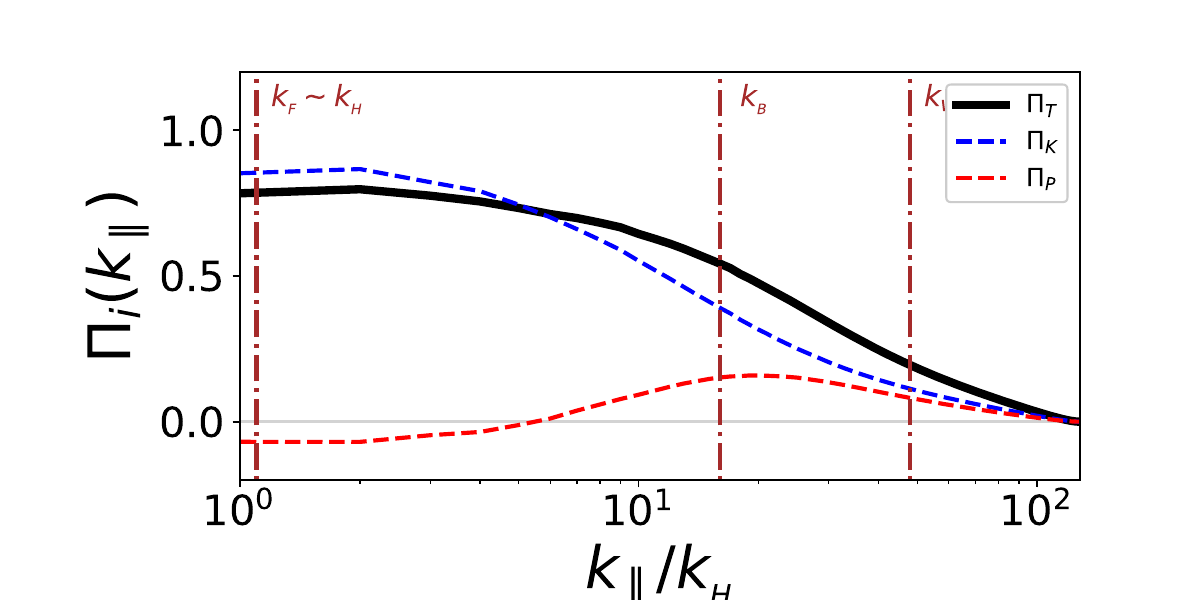} 
\caption{
{\bf Energy spectra and fluxes.} Left panels: Spherically (top), Cylindrically (middle), and plane (bottom) averaged energy spectra, for all energy components. 
Insets show the ratios of energy components $R_{GW}=E_{GW}/E_T$ (pink) and $R_{QG}=E_{QG}/E_T$ (green). 
Right panels: Energy fluxes across spheres (top), cylinders (middle), planes (bottom) in spectral space. Total energy flux (black line), energy flux of kinetic energy (blue line), energy flux of potential energy (red line). 
The forcing wavenumber $k_{_F} \approx \kH$ (where energy is injected),  the buoyancy wavenumber $k_B=N/U$, 
and the dissipation wavenumber $k_\nu$ (where energy is dissipated) are indicated by vertical dashed lines. 
The inset in the top right panel shows the inverse flux as a function of Re obtained from different runs.
\label{Fig:SpectraFluxes}
}
\end{figure*}                                        

The left panels in Fig. 2 
show the energy spectra $E_i(k)$, 
$E_i(k_\perp)$  and 
$E_i(k_\parallel)$  with the energy component $i$ as indicated in the legend.
The inset also shows the ratios $R_{GW}=E_{GW}/E_T$ and $R_{QG}=E_{QG}/E_T$.
In the top panel,
for $k>\kH$, the spectra have been averaged over shells of width $\kH$ since otherwise large peaks of period $\kH$ are observed due to the strong domain anisotropy, shown by the light grey lines for the total energy spectrum.
For wavenumbers larger than $\kH$ and smaller than the viscous wavenumber $k_\nu$
the spherically averaged spectrum displays a power-law behavior with an exponent very close to Kolmogorov's prediction $k^{-5/3}$ for homogeneous isotropic turbulence. This power-law behavior, composed 70\% by gravito-inertial waves,
is indicative of a forward energy cascade. 
At $k$ smaller than $\kH$ a similar power-law is observed (albeit with a smaller prefactor). This indicates the presence of an inverse cascade.
This energy at small $k$ is almost exclusively kinetic,  dominated by 
2D quasi-geostrophic modes.

%
For $E_i(k_\perp)$ three different power-laws can be observed. 
First, in the range $\kL<k_\perp<\kH$ a $k_\perp^{-5/3}$ law is observed, where $\kL = 2\pi/L$.
Second, in the range  $\kH<k_\perp<k_{_B}=N/U$ a steeper power-law close to $k_\perp^{-3}$ is observed,
where $k_{_B}$ is the buoyancy wavenumber.
Finally, at larger $k_\perp$, a shallower power-law slope starts to appear with 
exponent close to $-5/3$. 
The spectra for $k_\perp>\kH$ resemble the so-called Nastrom-Gage spectra \cite{nastrom1984kinetic} observed in the Earth's troposphere, although we should note here that, in the atmospheric case, the transition from the $k^{-3}$ regime to the $k^{-5/3}$ occurs at a scale $k'<\kH$
while here it occurs for $k>\kH$. 
%
Finally, the last panel of Fig. 2 
shows $E_i(k_\parallel)$, which displays a $k^{-5/3}$ power-law behavior. 

 The right panels of Fig. 2 
 show the energy
fluxes across different surfaces in wavenumber space: 
across constant $k$ spheres $\Pi_i(k)$,  constant $k_\perp$ cylinders $\Pi_i(k_\perp)$ and  constant $k_\parallel$ planes $\Pi_i(k_\parallel)$. 
As with the spectra, we distinguish between fluxes 
based on their arguments.
Here $i$ is $T$, $K$ or $P$ that stand for total, kinetic, and potential energies respectively.
Positive values imply a flux of energy towards large wavenumbers, while  negative values indicate a flux towards small wavenumbers.
%
$\Pi_T(k)$ flux is positive for $k>\kH$ indicating a forward cascade.
However, a small fraction, corresponding to 5\% of total energy injection rate, cascades towards the large scales.
This is seen in the negative flux observed at $k<\kH$. 
This flux is also constant for more than a decade of wavenumbers almost up to $\kL$. 
The inset in Fig.~2 
shows the amplitude of this negative flux measured from different simulations varying only $Re$. The flux  increases with $Re$ and saturates at the largest $Re$. 
$\Pi_T(k_\perp)$   is also positive for $k_\perp>\kH$ and negative for $k_\perp<\kH$. However, in this case the fraction of energy that cascades towards smaller $k_\perp$ is five times larger than $\Pi_T(k)$. 
 $\Pi_T(k_\parallel)$
is  positive everywhere. 
%
\begin{figure*}[!h]%
\centering
\includegraphics[width=0.48\textwidth]{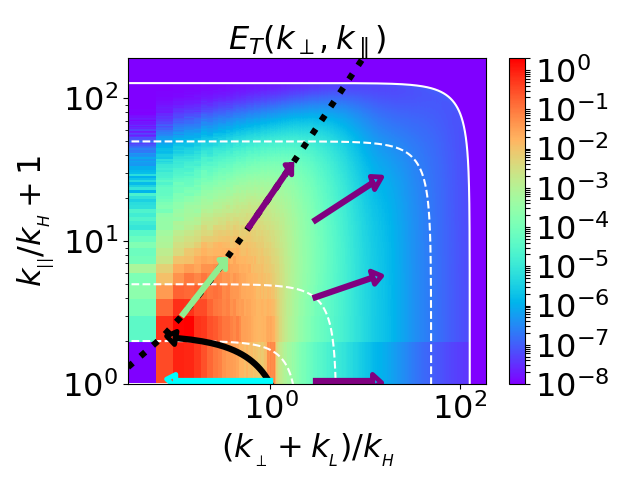}
\includegraphics[width=0.48\textwidth]{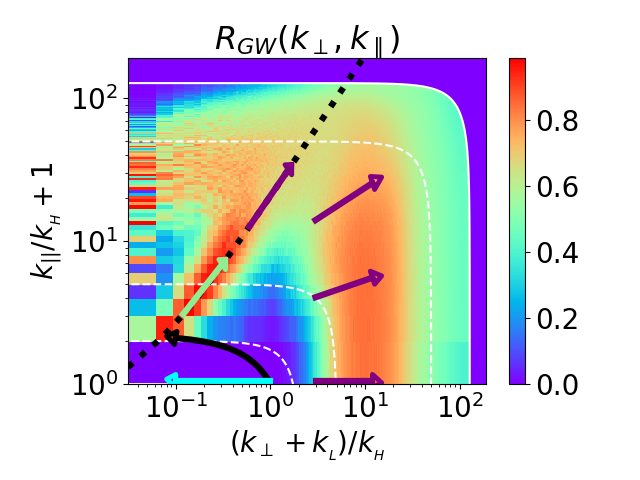}
\caption{{\bf Two-dimensional energy spectra in log-log scale.} White dashed lines indicate isotropic contours (i.e., modes with constant wavenumber $k$). The solid white line indicates the maximum resolved wavenumbers. Left: Total energy spectrum. Right: Gravity wave energy spectrum ratio. 
The black dotted lines marks $2\Omega k_\parallel = N k_\perp$ where inertial wave frequency matches gravity wave frequency.
The arrows indicate the direction of the flux of energy.
\label{fig:2D}}
\end{figure*} 

While in one-dimensional spectra and fluxes it is easier to identify power laws, the energy distribution depends 
on $k_\parallel$ and $k_\perp$ independently. 
In Fig.~3 
we show colour shaded plots of  $E_T(k_\perp,k_\parallel)$ and  $R_{GW}=E_{GW}(k_\perp,k_\parallel)/E_{T}(k_\perp,k_\parallel)$.
The arrows indicate the direction of the energy transfers based on the fluxes in Fig. 2. 
A part of the injected energy is transferred to larger $k_\perp$,  producing the $k_\perp^{-3}$ spectrum observed
in the $\kH <k_\perp<\kB$ range. $R_{GW}(k_\perp,k_\parallel)$ indicates that this forward transfer takes place through GW modes.
The peak of $E(k_\perp,k_\parallel)$ is observed at $k_\perp\simeq 2\kL$ and $k_\|\simeq 2\kH$
formed by an inverse transfer indicated by the black arrow and is dominated by QG modes.
This energy is responsible for the formation of the $k_\perp^{-5/3}$ scaling at small $k_\perp$ in Fig. 2. 
When rotation dominated scales are reached at $2\Omega k_\parallel \simeq N k_\perp $ (black dotted line),
QG modes transfer their energy to GW modes that cascade it back to small scales. 
Of the energy that has moved to smaller $k_\perp$, a finite amount is transferred (cyan arrow) 
below the smallest dashed white line ($k_\perp^2+k_\|^2=k_{_H}^2$). 
The energy in these modes forms the $E(k) \propto k^{-5/3}$ spectrum for $k<\kH$.
This energy
is the only that escapes to the largest scales $k\to0$
and corresponds to a true inverse cascade.

\section{Conclusions}   

This work has shown that dry turbulent atmospheres modeled by the non-hydrostatic Boussinesq equations can lead to a bidirectional energy cascade.
%
%
The results showed that indeed there is a flux of energy directed to the small wavenumbers $k$, 
corresponding to 5\% of the total energy injection rate at the largest Reynolds number. 
This flux, albeit small, is shown to persist up to the largest scales of the system and is $Re$-independent as large values of $Re$ are reached.

Our analysis provides a detailed description of how energy is transferred across scales and between different modes. 
These transfers,  indicated by the arrows sketched in Fig.~3, 
summarise
the results in this work.
Stratification, rotation and the geometric constraint of finite  $H$ all play a role in the formation of this inverse cascade.
In physical terms, at the scale of the forcing stratification is dominant, 
constraining a large fraction of the energy to hydrostatic (QG) modes.
This leads to the formation of pancake structures, known in stratified turbulence \cite{billant2001self}, that move energy to smaller $k_\perp$ and larger $k_\parallel$. 
This process ceases at wavenumbers where stratification is comparable to rotation $Nk_\perp \propto 2\Omega k_\parallel$. 
Rotation, that tends to bi-dimentionalise the flow \cite{greenspan1968theory}, prevents larger $k_\parallel$ modes from appearing 
and energy is converted to gravito-inertial mode energy that cascades back to larger $k$. 
This is true for all $\bf k$ except for the $k_\|=0$ modes that are unaffected by rotation. 
These modes that follow two-dimensional dynamics cascade the energy to ever smaller $k_\perp$.
Their stability against three-dimensional perturbations is assured by rotation and the finiteness of $H$  
that leads to
the $k_\parallel=0$ modes being isolated \cite{Alexakis_2018}. 
As a result they cascade energy to ever smaller $k_\perp$ with no channel to return this energy back to the small scales.  

An important outcome from this picture is that energy fluxes obtained from horizontal averages can significantly 
overestimate the inverse energy flux. In the presently examined simulation $\Pi_T(k_\perp)$
was five times larger than the true inverse flux $\Pi_T(k)$. This important result limits the 
observational estimates  $\Pi_T(k)$.  
 Most present estimates of the inverse flux are based on averages of two-dimensional slices obtained by satellite images.
They thus contain no information on the fields variation in the third direction and as a result it is $\Pi_T(k_\perp)$ 
that is measured and not  $\Pi_T(k)$ that represents the true inverse flux.
 As a reference, and for comparisons with observations, we provide as supplementary material (see Fig.~S1 in \cite{sm}) spectra and third-order structure functions along horizontal tracks, such as those resulting from airplane or satellite measurements. The structure functions display a change in sign indicative of an inverse cascade, but overestimate the inverse flux just as horizontal averages do.
 Thus these estimates of the inverse flux could be significantly larger than their true value.


 
%
%
%

\newpage

\section{Supplementary material}
\subsection*{Materials and Methods}

{}\hskip 0.6cm The Boussinesq equations (1) and (2) in the main text describe a rotating and stably stratified non-hydrostatic incompressible fluid. The pressure $P$ provides the correction to the hydrostatic pressure. The field ${\bf u}=(u,v,w)$ is the full three-dimensional velocity field, and the pointwise fluid density is $\rho = \rho_0 + z (d\bar{\rho}/d z) + \rho'$, where $\rho_0$ is the mean fluid density, $d\bar{\rho}/d z$ is the uniform background density gradient, and the density fluctuations $\rho'$ are written in terms of the normalised density variation $\phi$ as $\rho' = [-(\rho_0/g)(d\bar{\rho}/dz)]^{1/2} \phi$, where $g$ is the acceleration of gravity. 
Here we consider the case of stable stratification, $d\bar{\rho}/d z<0$, which is the case for a large part of the Earth's atmosphere. The squared Brunt-V\"{a}is\"{a}l\"{a} frequency is then $N^2 = -(g/\rho_0) (d\bar \rho /dz)$.
For large $N$ and in the presence of rotation we expect to approach the quasi-geostrophic limit, where gravity is balanced by pressure variations and vertical velocities and gravity waves are suppressed. In some regions of the atmosphere 
the stratification can be unstable, $d\bar{\rho}/d z>0$, which has been the focus of many studies \cite{Roberts_1968, Busse_1970, rubio2014upscale, favier2014inverse, Siegelman_2022}. In this case and for strong rotation a quasi-geostrophic limit can also be defined \cite{rubio2014upscale, favier2014inverse, Siegelman_2022} and even develop inverse energy cascades. It is worth noting that in the unstably stratified quasi-geostrophic regime gravitational modes are not entirely filtered out.

Simulations were done using the GHOST code, which is publicly available at 
\cite{GHOSTrepo}. Equations are integrated at double precision, using a parallel pseudo-spectral method to compute spatial derivatives, and a second-order Runge-Kutta method to evolve the fields in time \cite{Mininni_2011}. Aliasing instabilities were controlled using the $2/3$ rule. No subgrid model or effective transport parameters were used, and all relevant spatial and temporal scales in the system were explicitly resolved.

The forcing was delta-correlated in time, acting only on the horizontal components of the velocity ($u$ and $v$), and on all modes in the Fourier shell $k \in [0.9, 1.1] k_H$. To initialise the flow, simulations were done at increasing spatial resolutions of $1536^2 \times 48$, $3072^2 \times 96$, $6144^2 \times 192$, and $12288^2 \times 384$ grid points. In the simulations with $1536^2 \times 48$ and with $3072^2 \times 96$ grid points the flow was started from rest, and the simulations were spun up for $\approx 70$ forcing-scale turnover times, $\tau_f = (\epsilon k_H^2)^{-1/3}$, which was sufficient to display a well developed inverse cascade. The simulation at $6144^2 \times 192$ resolution was then done starting from the re-scaled last output of the previous simulation, running for at least $\approx 25$ turnover times. Finally, the simulation using $12288^2 \times 384$ grid points was integrated for $\approx 6$ turnover times from the last re-scaled state of the $6144^2 \times 192$ run. This time was sufficient for any transients to die out, and for a clear increase of energy due to the inverse cascade to be observed. To ensure the flows were spatially resolved we also performed other simulations at lower resolution using different Reynolds numbers. To remove initial transients, all analyses were performed using the last turnover times of each simulation, averaging over time.

We also performed a simulation with a resolution of $1536^2 \times 48$ grid points with no-slip boundary conditions at $z=0$ and $z=H$ using a Fourier continuation method \cite{Fontana_2020}, to verify that the inverse cascade also develops with different boundary conditions in the vertical direction. Finally, note that all simulations were stopped at a time such that energy did not accumulate at the lowest horizontal wave numbers, thus avoiding the formation of domain-scale features that would be impacted by the finite domain size.

Three-dimensional visualisations were done with the software VAPOR \cite{Clyne_2007}, publicly available at 
\cite{VAPORrepo}.

\vskip 24pt
\subsection*{Supplementary Text}

\noindent \underline{Energy decomposition}
\vskip 0.2 cm

The total energy per mass unit of the system, $\mathcal{E}_T$, is given by the sum of the kinetic and potential energies, given respectively by $\mathcal{E}_{K}=\langle \lvert {\bf u} \rvert ^2\rangle/2$ and $\mathcal{E}_{P}=\langle  \phi^2 \rangle/2$, where $\langle \cdot \rangle$ indicates the volume average over the domain. Alternatively, the energy can be split into the energy contained in gravity wave modes and in quasi-geostrophic modes. Following \cite{bartello1995geostrophic, herbert2014restricted}, we write for each wave number $\bf k$ 
the modes in terms of 
the vertical vorticity $\hat{w}_\parallel({\bf k})$, 
the vertical velocity  $\hat{u}_\parallel({\bf k})$, and 
the density variation  $\hat{\phi}({\bf k})$ as 
\[ 
{\bf Z}({\bf k}) = [ \hat{w}_\parallel({\bf k}),\, -ik\hat{u}_\parallel({\bf k}),\, -k_\perp \hat{\phi}({\bf k})] .
\] 
Assuming a periodic time dependence $e^{i \sigma t}$ and diagonalizing the linear part of the Boussinesq equations, we obtain three eigenfrequencies
\[
\sigma_{_{QG}}({\bf k})=0, \quad \mathrm{and} \quad 
\sigma^\pm_{_{GW}}({\bf k})=\pm \sqrt{(4\Omega^2k_\parallel + N^2 k_\perp^2)/k^2} ,
\] 
with $\sigma_{_{QG}}$ corresponding to the quasi-geostrophic mode and the other two to the gravity wave modes.
The eigenvectors are given by 
\[
{\bf Z}_{QG}({\bf k}) = [ iNk_\perp,\, 0,\, 2\Omega k_\parallel], 
\quad
\mathrm{and}
\quad 
{\bf Z}^\pm_{GW}({\bf k}) = [ i 2\Omega k_\parallel,\, -k\sigma^\pm({\bf k}),\, Nk_\perp].
\]
The projection $\mathcal{P}({\bf Z})$ of a mode $\bf Z$ to the quasi-geostrophic modes is given by
\[ 
\mathcal{P}_{QG}({\bf Z})={\bf Z}_{QG} [{\bf Z}_{QG}^*\cdot {\bf Z}]/\vert {\bf Z}_{QG}\vert^2, \] 
and for the gravito-inertial modes by
\[
\mathcal{P}_{GW}({\bf Z})=
{\bf Z}^+_{GW} \frac{({\bf Z}^+_{GW})^*\cdot {\bf Z}}{\vert {\bf Z}^+_{GW}\vert^2}+
{\bf Z}^-_{GW} \frac{({\bf Z}^-_{GW})^*\cdot {\bf Z}}{\vert {\bf Z}^-_{GW}\vert^2} .
\]
Energies and energy spectra can then be calculated using these projections. 

\vskip 0.5 cm
\noindent \underline{Energy spectra and fluxes}
\vskip 0.2 cm

Energy is distributed across scales by nonlinear interactions. 
The simplest way to quantify the distribution of energy among scales is by the Fourier transformed fields $\hat{\bf u}({\bf k})$ and $\hat{\phi}({\bf k})$, where the notion of scale is given by the inverse wavenumber $k^{-1}$.
The  kinetic and potential energy spectra are then defined respectively as 
\[
E_K(k_i)= \frac{1}{k_{_L}} \sum_{{\bf q}\in S_k} \vert \hat{\bf u}({\bf q})\vert^2, \qquad
E_P(k_i)= \frac{1}{k_{_L}} \sum_{{\bf q}\in S_k} \vert \hat{\phi}({\bf q}) \vert^2,
\]
with $k_{_L}=2\pi/L$, where $S_k$ is the set of wavenumbers $\bf q$ that satisfy $k \le \vert {\bf q} \vert < k+k_{_L} $ for the spherically averaged spectra, $k_\perp \le \vert {\bf q}_\perp \vert < k_\perp+k_{_L} $ for the cylindrically averaged spectra, and $k_\parallel \le \vert {q}_\parallel \vert < k_\parallel +k_{_L} $ for the plane averaged spectra.
For the two dimensional spectra, $S_k$ is the set that satisfies both $k_\perp \le \vert {\bf q}_\perp \vert < k_\perp +k_{_L} $ and $k_\parallel \le \vert {q}_\parallel \vert < k_\parallel+k_{_L} $.
The total energy spectrum is always given by $E_T(k)=E_K(k)+E_P(k)$.

To construct the spectra of gravito-inertial modes and of quasi-geostrophic modes, the same process is repeated after the modes $\hat{\bf u}({\bf k})$ and $\hat{\phi}({\bf k})$ are projected to the relevant manifolds as described before.

The kinetic and potential energy fluxes $\Pi_K(k)$ and $\Pi_P(k)$ through a closed set of wavenumbers $S_k$ are defined as
\[ 
\Pi_K(k) = \langle {\bf u}_{S_k} \cdot ({\bf u} \cdot \boldsymbol{\nabla} {\bf u}) \rangle, \qquad  \Pi_P(k) = \langle {\phi}_{S_k} ({\bf u}  \cdot \nabla \phi) \rangle ,
\]
where ${\bf u}_{S_k}$ and ${\phi}_{S_k}$ are obtained by filtering the fields ${\bf u}$ and $\phi$ so that only wavenumbers in the set $S_k$ are kept. 
The total energy flux is then given by $\Pi_T=\Pi_K+\Pi_P$. 
For the flux through a sphere the set $S_k$ corresponds to the set of wavenumbers ${\bf q}$ that satisfy $\vert {\bf q} \vert \le k$,
for the flux through cylinders it corresponds to the set of wavenumbers ${\bf q}$ that satisfy $\vert {\bf q}_\perp \vert \le k_\perp$,
and for flux through planes it corresponds to the set of wavenumbers ${\bf q}$ that satisfy $\vert q_\parallel \vert \le k_\parallel$. 

Note that the averages over all these manifolds are defined in spectral space. Thus, spectra and fluxes that depend on $k_\parallel$ quantify vertical gradients and structures in real space. Spectra and fluxes that depend on $k_\perp$ quantify horizontal gradients (as those obtained in atmospheric measurement from flights). Isotropic spectra and fluxes (i.e., dependent on $k$) mix gradients in all directions, and should be interpreted with care in the presence of anisotropies. However, the latter are the relevant spectral quantities to uniquely quantify cascades from a theoretical point of view.

\vskip 0.5 cm
\noindent \underline{Velocity spectra and structure functions along vertical and horizontal tracks}
\vskip 0.2 cm

We extracted 384 random horizontal tracks from the numerical data (to mimic, e.g., observations done along straight flight legs by airplanes, or reconstructions of velocity fields using velocimetry techniques from satellite images). We computed wind velocity ($u$ component) power spectra along each individual track, and third-order structure functions
\[
S_3(r) = \left< [u({\bf x} + r {\bf e}_x) - u({\bf x})]^3 \right>,
\]
where ${\bf e}_x$ is the unit vector along the track, and the average is done over all points ${\bf x}$ in the horizontal track. This structure function is expected to follow a power law and to be negative at scales in which energy is transfered to smaller scales, and positive when the energy flux is reversed \cite{Cho_2001, Lindborg_2001}.

Figure S1 shows the resulting spectra and structure functions along the synthetic tracks. Note the change in sign of the average of $S_3$ over all tracks at $k_H r \approx 1$, as well as the strong fluctuations in the individual tracks. Assuming $S_3(r) = -(4/5) \epsilon r$ \cite{Lindborg_2001}, the analysis indicates that the energy flux at large scales is negative (i.e., inverse), and results in an estimation of the inverse flux similar to the one obtained from $\Pi_T(k_\perp)$ (i.e., on an overestimation of the inverse flux). The estimation can be slightly improved by using scaling laws that take into account the flow stratification and anisotropy \cite{Lindborg_2001}, but as shown in \cite{Nie_1999}, in the anisotropic case proper determination of energy fluxes from $S_3$ requires averaging along multiple directions.

Comparisons with observations should take into account the difficulty in obtaining airplane data along long legs, and the dependence of satellite data on indirect methods to reconstruct velocity fields. In spite of these limitations, changes in the scaling law in airplane data \cite{Trent_2018} and positive third-order structure functions in airplane and satellite data \cite{Cho_2001, Heas_2011, Imazio_2022} have been reported. The scale at which scaling laws or the sign of $S_3$ change in these observations depends on atmospheric conditions, varying also across different samples and types of observations. While the results presented here can provide a possible explanation for this behavior, care must be taken to also consider fluctuations and noise in the data, the already mentioned overestimation of the inverse energy flux resulting solely from horizontal measurements, and sources of energy at large atmospheric scales that can obfuscate inverse fluxes.

\newpage

{}
\begin{figure}[h!]
\begin{center}
    \includegraphics[width=12cm]{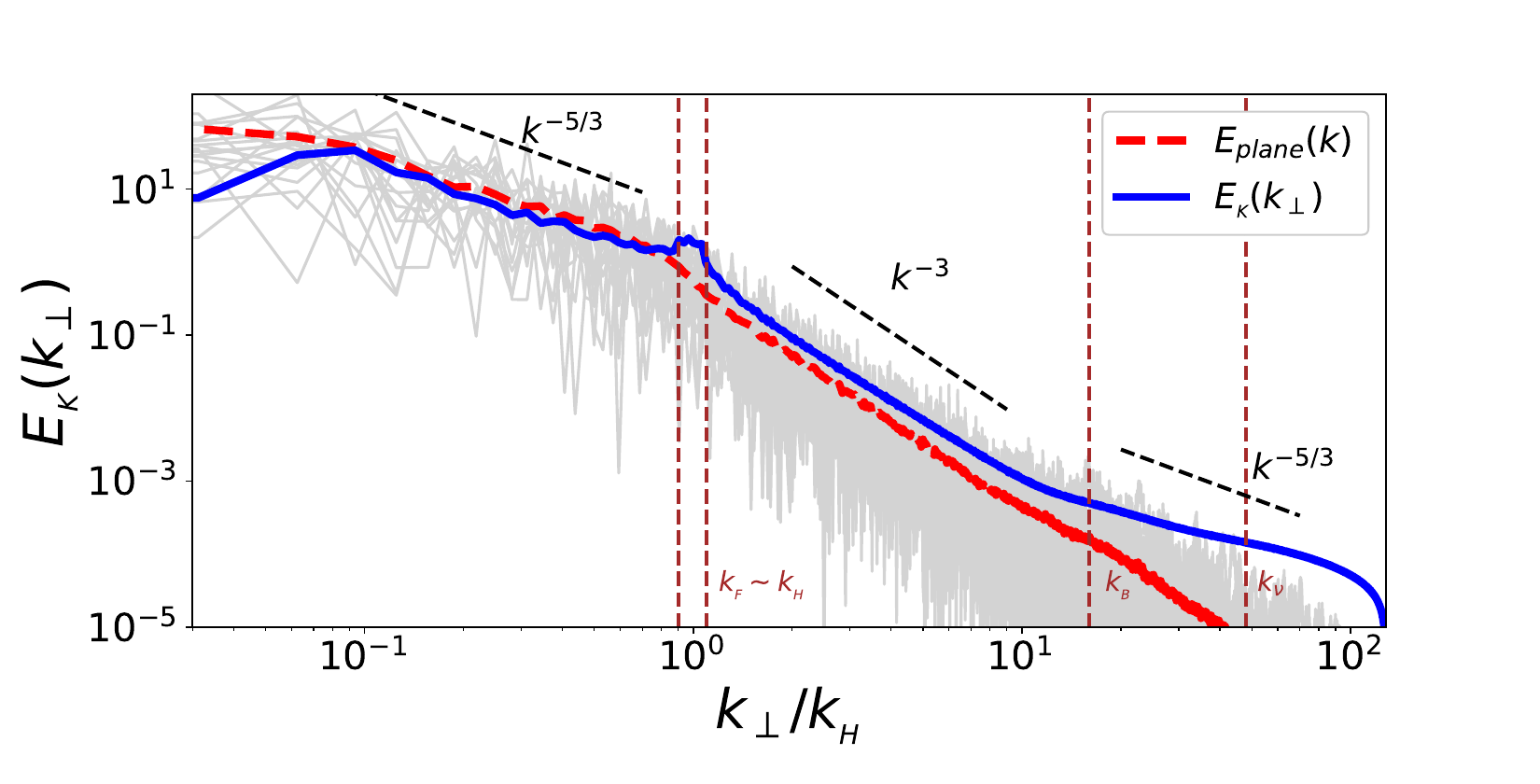}
    \includegraphics[width=12cm]{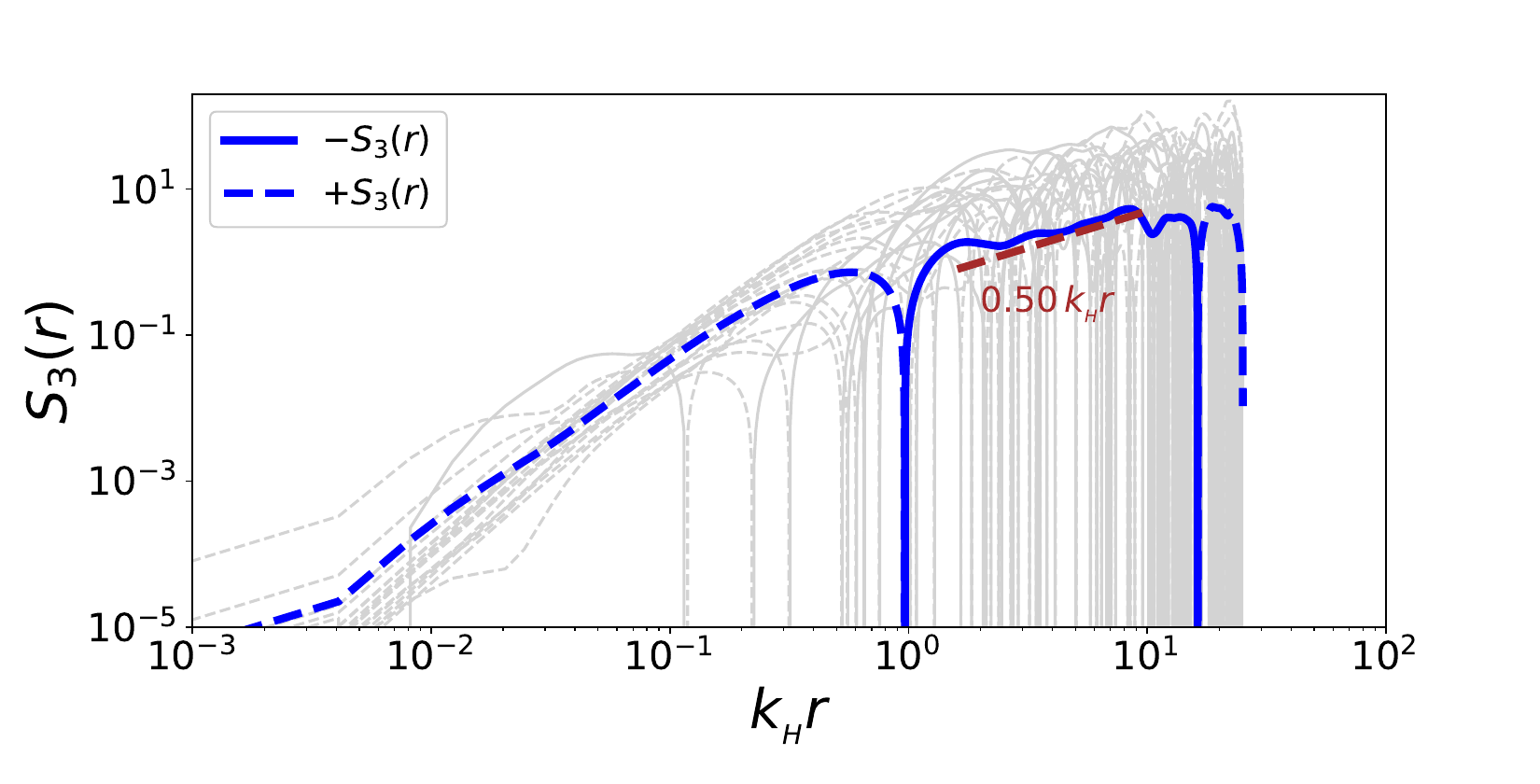}
\end{center}
\caption{{\bf Power spectrum and third-order structure function along synthetic tracks.} Top: Power spectrum of the $u$ component of the velocity field averaged along horizontal tracks in the $x$ direction (red line). The power spectrum obtained considering all directions and points in the simulation is shown in blue line. Gray lines show several spectra obtained for individual tracks, to illustrate variability. Characteristic scales and power laws are shown as a reference. Bottom: Third-order structure function averaged over horizontal tracks (solid and dashed, depending on the sign). Several structure functions for individual tracks are also shown in gray to illustrate variability. A power law is indicated as a reference in the range with negative energy flux.}
\end{figure}

\newpage

\noindent {\bf Movie S1. (separate file)}\\
A movie showing density fluctuations at different heights in the entire domain is provided as Supplementary Material. The movie features the development of large-scale structures as well as the formation of small scales. Note the emergence of macroscopic cyclones and anti-cyclones with relative lower and higer densities, as well as small-scale overturning events reminiscent of Kelvin-Helmholtz billows.




\bibliography{BIG_Turbulence_HAL}

\begin{thebibliography}{10}

\bibitem{Onsager1949}
L.~Onsager, {\it Il Nuovo Cimento (1943-1954)\/} {\bf 6}, 279 (1949).

\bibitem{Kraichnan1967}
R.~H. Kraichnan, {\it Physics of {F}luids\/} {\bf 10}, 1417 (1967).

\bibitem{Kraichnan1980}
R.~H. Kraichnan, D.~Montgomery, {\it Reports on Progress in Physics\/} {\bf
  43}, 547 (1980).

\bibitem{leith1968diffusion}
C.~E. Leith, {\it Physics of Fluids\/} {\bf 11}, 671 (1968).

\bibitem{batchelor1969computation}
G.~K. Batchelor, {\it Physics of Fluids\/} {\bf 12}, II (1969).

\bibitem{charney1971geostrophic}
J.~G. Charney, {\it Journal of the {A}tmospheric {S}ciences\/} {\bf 28}, 1087
  (1971).

\bibitem{Roberts_1968}
P.~H. Roberts, {\it Philosophical Transactions of the Royal Society of London.
  Series A, Mathematical and Physical Sciences\/} {\bf 263}, 93.

\bibitem{Busse_1970}
F.~H. Busse, {\it Journal of Fluid Mechanics\/} {\bf 44}, 441 (1970).

\bibitem{rubio2014upscale}
A.~M. Rubio, K.~Julien, E.~Knobloch, J.~B. Weiss, {\it Physical Review
  Letters\/} {\bf 112}, 144501 (2014).

\bibitem{favier2014inverse}
B.~Favier, L.~J. Silvers, M.~R.~E. Proctor, {\it Physics of Fluids\/} {\bf 26},
  096605 (2014).

\bibitem{Siegelman_2022}
L.~Siegelman, {\it et~al.\/}, {\it Nature Physics\/} {\bf 18}, 357 (2022).

\bibitem{vanneste2004exponentially}
J.~Vanneste, I.~Yavneh, {\it Journal of the {A}tmospheric {S}ciences\/} {\bf
  61}, 211 (2004).

\bibitem{vanneste2013balance}
J.~Vanneste, {\it {A}nnual {R}eview of {F}luid {M}echanics\/} {\bf 45}, 147
  (2013).

\bibitem{thomas2021forward}
J.~Thomas, D.~Daniel, {\it Journal of {F}luid {M}echanics\/} {\bf 911}, A60
  (2021).

\bibitem{thomas2020turbulent}
J.~Thomas, D.~Daniel, {\it Journal of Fluid Mechanics\/} {\bf 902}, A7 (2020).

\bibitem{Alexakis_2018}
A.~Alexakis, L.~Biferale, {\it Physics Reports\/} {\bf 767}, 1 (2018).

\bibitem{smith1996crossover}
L.~M. Smith, J.~R. Chasnov, F.~Waleffe, {\it Physical {R}eview {L}etters\/}
  {\bf 77}, 2467 (1996).

\bibitem{Celani_2010}
A.~Celani, S.~Musacchio, D.~Vincenzi, {\it Physical {R}eview {L}etters\/} {\bf
  104}, 184506 (2010).

\bibitem{seshasayanan2014edge}
K.~Seshasayanan, S.~J. Benavides, A.~Alexakis, {\it Physical {R}eview E\/} {\bf
  90}, 051003 (2014).

\bibitem{benavides2017critical}
S.~J. Benavides, A.~Alexakis, {\it Journal of {F}luid {M}echanics\/} {\bf 822},
  364 (2017).

\bibitem{sozza2015dimensional}
A.~Sozza, G.~Boffetta, P.~Muratore-Ginanneschi, S.~Musacchio, {\it Physics of
  Fluids\/} {\bf 27}, 035112 (2015).

\bibitem{van2020critical}
A.~van Kan, A.~Alexakis, {\it Journal of {F}luid {M}echanics\/} {\bf 899}, A33
  (2020).

\bibitem{marino2013inverse}
R.~Marino, P.~D. Mininni, D.~Rosenberg, A.~Pouquet, {\it EPL ({E}urophysics
  {L}etters)\/} {\bf 102}, 44006 (2013).

\bibitem{pouquet2013geophysical}
A.~Pouquet, R.~Marino, {\it Physical {R}eview {L}etters\/} {\bf 111}, 234501
  (2013).

\bibitem{Marino2015}
R.~Marino, A.~Pouquet, D.~Rosenberg, {\it Physical Review Letters\/} {\bf 114},
  114504 (2015).

\bibitem{byrne2013height}
D.~Byrne, J.~A. Zhang, {\it Geophysical {R}esearch {L}etters\/} {\bf 40}, 1439
  (2013).

\bibitem{king2015upscale}
G.~P. King, J.~Vogelzang, A.~Stoffelen, {\it Journal of {G}eophysical
  {R}esearch: {O}ceans\/} {\bf 120}, 346 (2015).

\bibitem{balwada_22}
D.~Balwada, J.-H. Xie, R.~Marino, F.~Feraco, {\it Science {A}dvances\/} {\bf
  8}, eabq2566 (2022).

\bibitem{young2017forward}
R.~Young, P.~L. Read, {\it Nature {P}hysics\/} {\bf 13}, 1135 (2017).

\bibitem{sm}
Materials and methods are available as supplementary materials.

\bibitem{pedlosky1987geophysical}
J.~Pedlosky, {\it Geophysical {F}luid {D}ynamics\/} (Springer, 1987).

\bibitem{vallis2017atmospheric}
G.~K. Vallis, {\it Atmospheric and {O}ceanic {F}luid {D}ynamics\/} (Cambridge
  University Press, 2017).

\bibitem{Mininni_2011}
P.~D. Mininni, D.~Rosenberg, R.~Reddy, A.~Pouquet, {\it Parallel Computing\/}
  {\bf 37}, 316 (2011).

\bibitem{hanli}
H.-L. Liu, P.~B. Hays, R.~G. Roble, {\it Journal of the Atmospheric Sciences\/}
  {\bf 56}, 2152  (1999).

\bibitem{Clyne_2007}
J.~Clyne, P.~Mininni, A.~Norton, M.~Rast, {\it New Journal of Physics\/} {\bf
  9}, 301 (2007).

\bibitem{nastrom1984kinetic}
G.~Nastrom, K.~Gage, W.~Jasperson, {\it Nature\/} {\bf 310}, 36 (1984).

\bibitem{billant2001self}
P.~Billant, J.-M. Chomaz, {\it Physics of Fluids\/} {\bf 13}, 1645 (2001).

\bibitem{greenspan1968theory}
H.~P. Greenspan, {\it The theory of rotating fluids\/} (Cambridge University
  Press, 1968).

\bibitem{GHOSTrepo}
D.~Rosenberg, P.~Mininni, P.~C.~D. Leoni, pmininni/ghost: Ghost master, version
  2.0.0, zenodo (2023); https://doi.org/10.5281/zenodo.8015308.

\bibitem{Fontana_2020}
M.~Fontana, O.~P. Bruno, P.~D. Mininni, P.~Dmitruk, {\it Computer Physics
  Communications\/} {\bf 256}, 107482 (2020).

\bibitem{VAPORrepo}
S.~e.~a. Pearse, Ncar/vapor: Vapor, version 3.8.1, zenodo (2023);
  https://doi.org/10.5281/zenodo.7779648.

\bibitem{bartello1995geostrophic}
P.~Bartello, {\it Journal of {A}tmospheric {S}ciences\/} {\bf 52}, 4410 (1995).

\bibitem{herbert2014restricted}
C.~Herbert, A.~Pouquet, R.~Marino, {\it Journal of {F}luid {M}echanics\/} {\bf
  758}, 374 (2014).

\bibitem{Cho_2001}
J.~Y. Cho, E.~Lindborg, {\it Journal of Geophysical Research: Atmospheres\/}
  {\bf 106}, 10223 (2001).

\bibitem{Lindborg_2001}
E.~Lindborg, J.~Y. Cho, {\it Journal of Geophysical Research: Atmospheres\/}
  {\bf 106}, 10233 (2001).

\bibitem{Nie_1999}
Q.~Nie, S.~Tanveer, {\it Proceedings of the Royal Society of London. Series A:
  Mathematical, Physical and Engineering Sciences\/} {\bf 455}, 1615 (1999).

\bibitem{Trent_2018}
P.~{Trent Vonich}, G.~J. Hakim, {\it Journal of the Atmospheric Sciences\/}
  {\bf 75}, 2523  (2018).

\bibitem{Heas_2011}
P.~H{\'e}as, E.~M{\'e}min, D.~Heitz, P.~D. Mininni, {\it Tellus A: Dynamic
  Meteorology and Oceanography\/} {\bf 64}, 10962 (2011).

\bibitem{Imazio_2022}
P.~Rodriguez~Imazio, A.~D{\"o}rnbrack, R.~D. Urzua, N.~Rivaben, A.~Godoy, {\it
  Journal of Geophysical Research: Atmospheres\/} {\bf 127}, e2021JD035908
  (2022).

\end{thebibliography}



\bibliographystyle{Science}

\section*{Acknowledgements}


{\bf Funding:} Computer resources in Joliot-Curie at CEA were provided by PRACE (research project ID 2020235566), and by GENCI (allocation No.~A0110506421). This work was supported by the projects ``DYSTURB" (No.~ANR-17-CE30-0004) and ``EVENTFUL" (No.~ANR-20-CE30-0011) funded by the French ``Agence Nationale de la Recherche" - ANR, by the Studienstiftung des deutschen Volkes, by the National Science Foundation (Grants DMS-2009563, DMS-2308337) and by the German Research Foundation (Projektnummer: 522026592).  The simulation output was analyzed on HPC facilities at the École Normale Superieure in Paris (France), at École Centrale de Lyon (PMCS2I) in Ecully (France), and at Departamento de Fisica (FCEN, UBA) in Argentina.






\end{document}